\documentclass{ws-ijgmmp}
\usepackage{color}

\begin{document}

\markboth{F. Darabi, K. Atazadeh}
{Einstein static universe, GUP, and natural IR and UV cut-offs}

%
\catchline{}{}{}{}{}
%

\title{Einstein static universe, GUP, and natural IR and UV cut-offs}

\author{F. Darabi\footnote{Corresponding author}}

\address{Department of Physics, Azarbaijan Shahid Madani University, Tabriz, 53714-161 Iran.\\
              \email{f.darabi@azaruniv.ac.ir} }

\author{K. Atazadeh}

\address{Department of Physics, Azarbaijan Shahid Madani University, Tabriz, 53714-161 Iran.\\
              \email{atazadeh@azaruniv.ac.ir} }

\maketitle

\begin{history}
\received{(Day Month Year)}
\revised{(Day Month Year)}
\end{history}

\begin{abstract}
We study the Einstein static universe in the framework of Generalized Uncertainty Principle constructed by the Snyder non-commutative space. It is shown that the deformation parameter can induce an effective energy density subject to GUP which obeys the holographic principle (HP) and plays the role of a cosmological constant. Using the holographic feature of this effective energy density, we introduce natural IR and UV cut-offs which depend on the GUP based effective equation of state. Moreover, we propose a solution to the cosmological constant problem. This solution is based on the result that the Einstein equations just couple to the tiny holographic based surface energy density (cosmological constant) induced by the deformation parameter, rather than the large quantum gravitational based volume  energy density (vacuum energy) having contributions of order $M_P^4$.
\end{abstract}



\section{Introduction}

In the last decade, a new model of dark energy (DE)  so called Holographic Dark Energy (HDE) was proposed
based on the holographic principle (HP)
\cite{cohen,li}. The energy density for  HDE  is given by \cite{susskind}-\cite{susskind2}
\begin{equation}\label{1}
\rho_{\Lambda} = 3 c^{2} M_{p}^{2} L^{-2},
\end{equation}
where $c$ is a numerical constant, $M_{p}$ is the  Planck mass and $L$
is the infrared IR cut-off length. Different choices for IR cut-off length
have been proposed such as Hubble length, particle horizon, event
horizon, apparent horizon, Ricci radius, Granda-Oliveros cut-off, and so on \cite{N-O}.  On the other hand, the existence of a minimal length  is a prediction of quantum theory of gravity \cite{a}-\cite{f}.  Therefore, at high energy physics  such as early universe  we must consider the effects of such a minimal length.
Such consideration is achieved by
the deformation of  standard Heisenberg commutation relation known as the Generalized Uncertainty Principle (GUP) {\cite{g}-\cite{capo}}. The simplest form of such relation in the framework of  Snyder non-commutative space
is given by \cite{Snyder}
\begin{equation}\label{uncrel}
\Delta q \Delta p \geq \frac{1}{2} |<\sqrt{1 - \alpha p^2}>|,
\end{equation}
{where we have used the units $\hbar=1$. In this units, $\alpha$ is a deformation parameter, with the dimension of squared length ($Length^2$), which is assumed to have a sufficiently small value. In fact, it is treated as a small parameter with $\sqrt{\alpha}$ being the measure of a minimum length scale \cite{Pram}. { The important point which is worth to mention here is that GUP is applicable to all energy regimes, however the question that at which energy scale it becomes
more important and considerable depends on the deformation parameter $\alpha$.}
The smallness of $\alpha$ is considered in relation to the order of magnitude of the momentum $p$, so that at low energy regime with classical values of momentum $p$, much smaller than the Plank energy, the correction term almost vanishes and we get the Heisenberg Uncertainty Principle, whereas at high energy regime with ultra-relativistic values of momentum $p$, of the order of Plank energy, the correction term becomes relevant and we get the Generalized Uncertainty Principle. The corresponding commutation relation can be written as}

\begin{equation}\label{xp}
[ q, p ] = i \sqrt{ 1 - \alpha p^2 },
\end{equation}
where the only freedom is on the sign of the deformation parameter $\alpha$. Therefore, a maximum momentum or a minimal length are predicted by the Snyder-deformed relation (\ref{xp}) if $\alpha>0$ or $\alpha<0$, respectively.

On the other
hand, the emergent ``Einstein Static Universe'' (ESU)
scenario was proposed by Ellis {\it et al} to solve the initial singularity problem in the standard cosmological model \cite{8}, \cite{81}. { Initial singularity is the major problem of most cosmological
models constructed based on General Relativity (GR), and there are variety of classical or quantum approaches to resolve this major problem. Recently, based on the loop quantum cosmology, the search for singularity free cosmological models within the framework of GR has led to development of the so-called ESU scenario \cite{8}, \cite{81}. In this scenario, the early universe is initially in a past-eternal ESU state and eventually evolves to a subsequent inflationary phase as follows. 

The stable ESU against cosmological perturbations addresses the existence of a fixed point around which our universe in the early times was eternally fluctuating. In the case of the universe filled by a massless scalar field with a suitable inflaton potential,
these fluctuations could have been disappeared and the stable state of the
universe might have been turned into an inflationary phase. This
transmission from a stable ESU to an inflationary phase could have been taken
place around the different values of e-folds. Of course, the
quantum field theory on curved spacetime and its modifications on the
Hilbert–Einstein action may affect deeply the inflationary scenarios \cite{Cappo}. Therefore, transmission from a stable ESU to an inflationary phase could have been also affected by these modifications. 

The achievements of this cosmological model are remarkable: there is no initial singularity; the universe is ever existing and it tends to a static universe in the past infinity rather than originating from a big bang singularity. Note that the emergent ESU scenario is completely different from that of introduced first by Einstein. The model of Einstein was introduced to describe a stable large scale universe, whereas the emergent ESU scenario which was introduced first by Ellis {\it et al} is related to the very early universe and the resolution of its problems such as initial singularity problem.
In fact, emergent ESU belongs to the very early era of the universe, even before the inflationary era. For this reason, the Einstein Static Universe ESU is considered to be related to the High Energy Regime of the early
universe.}
In this letter, we
study the impact of Generalized Uncertainty Principle (GUP) on the  Einstein Static Universe (ESU), Holographic Dark Energy (HDE) and  also the Cosmological Constant Problem (CCP), in the context of  Emergent Universe.

\section{GUP and ESU}

In this section, we investigate the consistency of the emergent Einstein Static Universe scenario with GUP. Let us consider the isotropic and homogeneous FRW universe described by the line element
\begin{equation}
d s^2 = - N^2 dt^2 + a^2 (\frac{dr^2}{1 - k r^2} + r^2 d \Omega^2),
\end{equation}
where $N=N(t)$ and $a=a(t)$ are the lapse function and  scale factor, respectively, and the spatial curvature $k$ can be zero or $\pm1$ depending on the symmetry group. We assume that this universe is filled by a perfect fluid given by
$T_{\mu\nu}=diag(\rho, p, p, p)$.
Hamiltonian constraint of such models is given by
\begin{equation}\label{scacon}
{\cal H}=-\frac{2\pi G}{3}\frac{p_{a}^{2}}{a}-\frac{3}{8\pi G}ak+a^{3}\rho=0,
\end{equation}
where $G$ is the gravitational constant, $\rho=\rho(a)$ denotes for generic energy density of the system and $p_{a}$ is the momentum conjugate to the scale factor $a$.
The dynamics of this model universe is described by the analysis of Hamiltonian constraint system which results in the following Poisson bracket equations
\begin{eqnarray}\label{eqap}
\dot N = \{N, \mathcal H_E\},~~~
\qquad \dot a = \{a,\mathcal H_E\},~~~
\qquad \dot p_a = \{p_a,\mathcal H_E\},
\end{eqnarray}
where
\begin{equation}\label{extham}
\mathcal H_E = \frac{2\pi G}{3 N} \frac{p_a^2}{a} + \frac{3}{8\pi G} N a k - N a^3 \rho
+ \lambda \pi.
\end{equation}
$\mathcal H_E$ is the extended Hamiltonian and  $\pi$ is the momenta conjugate to the lapse function. The Hamiltonian dynamics for cosmologies coming from extended theories of gravity have been considered in \cite{capo2}, \cite{capo3}.

By using  the above equations, based on the Poisson bracket $\{a,p_{a}\}=1$,
we can obtain the standard Friedmann equation.
Now, we  study the analysis of deformed dynamics of the FRW model to obtain the modifications of the  Friedmann equation resulting from the algebra (\ref{xp}).
The modified symplectic geometry at the classical limit of the algebra  (\ref{xp}), results in the Snyder-deformed classical dynamics. According to Dirac, we
can replace the quantum-mechanical commutator (\ref{xp}) by the classical
Poisson bracket
\begin{equation}\label{pm}
-i [\tilde q, p]\Longrightarrow\{\tilde q, p\} = \sqrt{1 - \alpha p^2}.
\end{equation}
  The deformed Poisson bracket must satisfy the same properties as those
of quantum mechanical commutator, i.e. it has to be bilinear and anti-symmetric,  and also it must satisfy the Leibniz rules as well as the Jacobi identity. Thus, the deformed Poisson bracket in  two-dimensional phase space is
given by\begin{equation}
\{ F, G\}=( \frac{\partial F}{\partial\tilde q} \frac{\partial G}{\partial p} - \frac{\partial F}{\partial p}\frac
{\partial G}{\partial\tilde q} ) \sqrt{1- \alpha p^2}.
\end{equation}
Specially, the canonical equations for coordinate and momentum from the deformed Hamiltonian $\mathcal H(\tilde q,p)$ are obtained
\begin{eqnarray}
\dot{ \tilde q}& = &\{\tilde q,\mathcal H\} = \frac{\partial\mathcal H}{\partial p} \sqrt{1-
\alpha p^2},\nonumber\\ \qquad \dot p& = &\{p,\mathcal H\} = -\frac{\partial \mathcal H}{\partial \tilde q} \sqrt{1 - \alpha p^2}.
\end{eqnarray}
Now, we apply this deformation scheme to the FRW model in the presence of  matter energy density, namely to the Hamiltonian (\ref{extham}). We assume the minisuperspace to  be Snyder-deformed and consequently the commutator between the scale factor $a$ and its conjugate momentum $p_a$ is uniquely obtained as
\begin{equation}\label{ap}
\{a, p_a\} = \sqrt{1- \alpha p_a^2}\,.
\end{equation}
{Considering the term $\alpha p_a^2$ in RHS, and the units $\hbar=1$, it is obvious that the dimension of $\sqrt{\alpha}$ is the same dimension of the conjugate coordinate of the momentum $p_a$, namely
the scale factor $a$. Therefore, since the dimension of $\sqrt{\alpha}$ is $Length$, then one may attribute the dimension of a $Length$ to the scale
factor too, as the radius of the universe\footnote{With this choice, the comoving radial coordinate $r$ becomes dimensionless \cite{a-r}.}.}

The Snyder-deformed commutation relation (\ref{ap}) does not change the equations of motion $\dot N=\{N,\mathcal H_E\}=\lambda$ and $\dot\pi=\{\pi,\mathcal H_E\}=\mathcal H=0$. The Poisson bracket $\{N,\pi\}=1$ is not affected by the deformations induced by the $\alpha$ parameter. Nevertheless, the equations of motion (\ref{eqap}) are modified in this approach via the relation (\ref{ap}), so we have
\begin{eqnarray}\label{eqapgup}
\dot a& =& \{a, \mathcal H_E\} = \frac{4 \pi G}{3 N} \frac{p_a}{a} \sqrt{1 - \alpha p_a^2}, \\ \nonumber
\qquad \dot p_a& = &\{p_a, \mathcal H_E\}= \sqrt{1 - \alpha p_a^2}
N (\frac{ 2 \pi G}{3} \frac{p_a^2}{a^2} - \frac{3}{8 \pi G} k + 3a^2 \rho + a^3
\frac{d\rho}{da} ).
\end{eqnarray}
The deformed Friedmann equation  can be obtained by solving the constraint (\ref{scacon}) with respect to $p_a$ and considering the first equation of (\ref{eqapgup}) as follows ($N=1$) \cite{marco}
\begin{equation}\label{deffri}
(\frac{\dot a}{a} )^2 = (\frac{ 8 \pi G}{3}\rho-\frac{k} {a^2} )[
1 - \frac{ 3\alpha}{2\pi G} a^2 ( a^2 \rho-\frac{3}{ 8 \pi G} k ) ].
\end{equation}
 The conservation equations for the matter part is
also given by
\begin{equation}\label{ma}
\dot{\rho} +3 H (\rho+p)= 0\,.
\end{equation}
By using the equations (\ref{deffri}) and (\ref{ma}) for the closed FRW universe ($k=1$) and the equation of state $p=w \rho$, the acceleration equation is obtained
\begin{eqnarray}\label{aa}
\ddot{a}& = &[-\frac{9 (w+1) \alpha  a^3}{2 \pi G}+\frac{6 \alpha  a^3}{ \pi G}+\frac{8 \pi G a}{3}-4 \pi G a (w+1)] \rho+12 w \alpha  \rho ^2 a^5-\frac{9 \alpha  a}{16 \pi^2 G^2 },
\end{eqnarray}
where the matter energy density $\rho$ is given by solving the equation (\ref{deffri}) as follows \cite{ata}
\begin{eqnarray}\label{bb}
\rho=\frac{9 G \pi  \alpha  a^4+8 G^3 \pi ^3 a^2 \pm 4 \sqrt{4 a^4 G^6 \pi ^6-9 a^8 G^4 H^2 \pi ^4 \alpha }}{24 a^6 G^2 \pi ^2 \alpha }.
\end{eqnarray}
Imposing the requirements of Einstein static solution given by  $\ddot{a} = \dot{a}=0 $ in
the  equations (\ref{aa}) and using (\ref{bb}), we obtain the solutions  $a_{_{\alpha}}(\omega)$ and the corresponding energy density $\rho_{_{\alpha}}(\omega)$ as follows
\begin{eqnarray}\label{20}
a^{2}_{_{\alpha}}(\omega)=\frac{16 \pi^2 G^2 }{9 \alpha}\frac{(1-3 w)}{(1+3 w)}~~,~~~~~~~\rho_{_{\alpha}}(\omega)=\frac{27 \alpha}{64 \pi^3 G^3  }\frac{(1+3 w)}{(1-3 w)^2},
\end{eqnarray}
where we call $\rho_{_{\alpha}}(\omega)$ as the effective GUP based energy
density. For a positive $\alpha$, the  reality condition for $a_{_{\alpha}}(\omega)$ and the positivity condition of $\rho_{_{\alpha}}(\omega)$ result in the following domain for the equation
of state parameter
\begin{eqnarray}\label{www}
-1/3<w<1/3.
\end{eqnarray}
By removing the parameter $\alpha$, the equation (\ref{20}) can be rewritten as
\begin{eqnarray}\label{20'}
~~~~~~\rho_{_{\alpha}}(\omega)=\frac{3}{4 \pi G (1-3w)a^{2}_{_{\alpha}}}.
\end{eqnarray}
{Although $\alpha$ is assumed to be a very small parameter, however the size of Einstein static universe $a_{_{\alpha}}(\omega)$ in (\ref{20})
can be sufficiently small, by appropriate choice of $\omega\lesssim \frac{1}{3}$, such that it can describe an emergent small size Einstein static universe with large energy density $\rho_{_{\alpha}}(\omega)$ at very early time. This means that such an emergent small universe is just consistent with the upper limit of the domain (\ref{www}), namely the ultra-relativistic state of the effective  matter $p_{_{\alpha}}(\omega)=w \rho_{_{\alpha}}(\omega)$ whose origin is GUP. This is in complete consistency with the origin of GUP described by (\ref{xp}) (or (\ref{ap})) at high energy regime with ultra-relativistic values of momentum $p$, of the order of Plank energy. In fact, the GUP correction term $\alpha p_a^2$ in (\ref{ap}) and the term ${(1-3 w)}/{\alpha}$ (or ${\alpha}/{(1-3 w)^2}$) in (\ref{20}) represent a similar and common
behavior that: in both cases the small parameter ${\alpha}$ is accompanied by the high energy state of matter, namely very large momentums $p_a$ or an ultra-relativistic state of matter with equation of state parameter $\omega\lesssim \frac{1}{3}$. Therefore, it turns out that not only the emergent Einstein Static Universe is consistent
with GUP, but also the implementation of GUP at high energy regime for the description of emergent Einstein Static Universe at very early universe is inevitable, and this certifies our motivation for the study of Einstein Static Universe in the framework of GUP.}

\section{Natural $w$ dependent IR-UV cut-offs }

{Comparing (\ref{20'}) with the generic form of the HDE, described
by the equation (\ref{1}) ($8\pi G=M_P^{-2}$), we find that the energy density $\rho_{_{\alpha}}(\omega)$ corresponding to the Einstein static universe has a holographic feature. Therefore, one may consider $L_{_{\alpha}}(\omega) = a_{_{Es}}(\omega)\sqrt{(1-3w)/2c^2 }$ as a $w$ dependent cut-off for the HDE. At upper limit $\omega\lesssim \frac{1}{3}$, which describes the ultra-relativistic state, this cut-off is essentially of UV type. At lower limit $\omega \gtrsim-\frac{1}{3}$, which describes the threshold of exotic matter, this cut-off is essentially of IR type due to the very large size of $a_{_{Es}}(\omega)$.
The dependence of IR-UV cut-offs on the equation of state parameter makes it a natural cut-off which is uniquely set by the natural state of material content\footnote{Relevant works to natural cut-offs are reported in Ref.\cite{Manos}}.

The extreme condition $\omega\lesssim \frac{1}{3}$ makes $\rho_{_{\alpha}}(\omega)\gg
0$, $a_{_{Es}}(\omega)\ll 1$ and causes the state of effective  matter to be ultra-relativistic
$p_{_{\alpha}}(\omega)\simeq \rho_{_{\alpha}}(\omega)/3$.
These are physical characteristic features of emergent Einstein Static Universe whose size is extremely ``small" and its matter content is extremely``dense" and ``hot". Such features can describe a universe which has capability of experiencing a Hot Big Bang, starting with an arbitrarily large energy density and an infinitesimally small size. However, since we have a finite and non-vanishing radius $a_{_{Es}}$, this Big Bang will be free of singularity problem. The singularity is avoided by resorting to the natural UV cut-off
imposed by the natural state of the GUP based matter content given by $\omega\lesssim \frac{1}{3}$.

\section{Discussion}

The equations of motion (\ref{deffri}) and (\ref{aa}) in the absence of $\alpha$ and without a cosmological constant $\Lambda$ lead to the usual Friedmann equations which describe an expanding universe. However, in the presence of $\alpha$
and without a cosmological constant  $\Lambda$,
we obtain Einstein static universe. This clearly indicates that in the absence of cosmological constant  $\Lambda$, the static state of  Einstein universe can be provided by the quantum gravitational effect induced by the deformation parameter $\alpha$. In other words, the deformation parameter $\alpha$ induces energy
densities $\rho_{_{\alpha}}$ and  $\rho_{_{\alpha}}(\omega)$
which  can play the role of  cosmological constants capable of constructing the Einstein static universe. These energy densities obey the holographic principle and ensures us about the fact that the origin of
these energy densities lie on the two dimensional surface of a sphere with the radius of Einstein static universe. The bounds on
$\alpha$ should be a subject of precise cosmological observations, for
instance the measurement of the cosmological constant.

For the case of $\rho_{_{\alpha}}$, by identification of $\rho_{_{\alpha}}$
with the cosmological constant we can estimate the value of  deformation parameter $\alpha$ by considering the observational upper bound on the present
value of cosmological
constant as $\alpha \sim (\Lambda\sim 10^{-47} GeV^4)$.
This also gives estimation on the size of the  radius of Einstein static universe as $a_{\alpha}=a_E\sim 10^{28}m$ which coincides with the size of
the observable universe. 

We may also study the cosmological constant problem in the  present model for the energy density $\rho_{_{\alpha}}$. The cosmological constant problem arises because of the huge
disagreement between the observed value of cosmological constant, of order $10^{-47} GeV^4$, and   the zero-point energy suggested by quantum field theory (QFT).  
We know that at high energies and small scales comparable to the Compton wavelength of the particles, matter is quantized and one can employ a semiclassical description of gravitation in which the Einstein equations take the form
$$
G_{\mu\nu}=\langle T_{\mu\nu}\rangle,
$$
where $G_{\mu\nu}$ is the Einstein tensor and $\langle T_{\mu\nu}\rangle$ denotes the expectation value
of a quantum stress-energy tensor. In curved spacetime, even in the absence of classical matter and radiation, quantum fluctuations of matter fields give non-vanishing contributions to $\langle T_{\mu\nu}\rangle$ through a non-minimal couplings between matter and the curvature \cite{Cappo}. Considering such
non-vanishing contributions, of order $M_{p}^{4}\sim 10^{74} GeV^4$, reveals a large discrepancy of order $10^{121}$ between the observed and calculated
values of zero point energy.

In principle, the zero-point energy in quantum field theory is  provided by the vacuum quantum field fluctuations originated by the Heisenberg uncertainty principle
$\Delta q \Delta p \geq \frac{1}{2}$. By applying (\ref{uncrel}), we expect that the leading ``unit'' term in the square root is responsible for the zero-point energy of quantum gravity which can be originated
by the vacuum fluctuations corresponding to Generalized Uncertainty Principle $\Delta q \Delta p \geq \frac{1}{2} |\langle \sqrt{1 - \alpha p^2} \rangle |$, and the small correction term $\alpha p^2\sim \alpha M_{p}^{2}\sim 10^{-10}\ll1$ has negligible
contribution to the zero-point energy of quantum gravity, hence it is not expected that the correction term $\alpha p^2$ can cancel out the huge zero-point energy of quantum gravity and solve the cosmological constant problem. However, the application of GUP on the Einstein static
universe has a capability of resolving the cosmological constant problem
in this model, as is described in the following.

We know that the zero-point energy
density is the zero-point energy per unit volume,
namely it is a  energy density distributed over the entire volume inside
the Einstein static universe, whereas the energy density  $\rho_{_{\alpha}}$ corresponding to the deformation parameter $\alpha$ is a surface energy density
distributed over the entire surface enclosing
the Einstein static universe.
The fact that in spite of the presence of quantum gravitational vacuum fluctuations of order $M_P^4$ distributed over the volume of Einstein static universe,
the cosmological constant in the Einstein equation (which arises effectively from GUP in the  Einstein static universe)
is merely set and established by the holographic surface energy density
$\rho_{_{\alpha}}$ distributed over the surface enclosing the Einstein static universe, may remark the important point that the holographic based cosmological constant $\Lambda \sim\rho_{_{\alpha}}$ with the origin over the enclosing surface of Einstein static universe is the real cosmological
constant which contributes to the Einstein equation, and the zero-point volume energy density raised by the quantum gravitational corrections of order $M_P^4$ distributed over the volume of Einstein static universe is an irrelevant
and nonphysical background energy which has vanishing contribution to the Einstein equation.
In simple words, in the sprit of holographic principle, it seems that the observed cosmological constant $ \Lambda \sim\rho_{_{\alpha}} $ has just a holographic nature which is merely provided
by the information encoded on the surface of the sphere enclosing the Einstein static universe, rather than the volume information inside the Einstein static universe. Therefore, one may conclude that the volume information including the quantum gravitational perturbative corrections, in principle cannot contribute to the surface information including the classical energy density  $\rho_{_{\alpha}}$ appearing in the Einstein equations. It turns out that, according
to holographic principle, the Einstein tensor just feels (and react
against) the small holographic surface
energy density of order $10^{-47} GeV^4$ rather than the huge volume energy density of order $10^{74} GeV^4$. Therefore, similar to the  quantum field theory  in flat spacetime where  the
zero of energy is arbitrary and the vacuum energy with infinite energy is considered as a nonphysical background energy,   the volume energy density here which is not felt by the Einstein tensor, can also be regarded as a nonphysical background energy, as explained above.

It is well-known that the zero-point energy is fundamentally related to the Heisenberg Uncertainty Principle stated as $\Delta q \Delta p \geq \frac {1}{2}$
or the commutation relation $[q, p] = i $. Since the zero-point energy is distributed over the entire space, it is obviously a volume energy density
which can be attributed to non-vanishing $``\frac{1}{2}"$ (or $``i"$) in  $\Delta q \Delta p \geq \frac{1}{2}$ (or $[q, p] = i $). Considering the Generalized Uncertainty Principle $\Delta q \Delta p \geq \frac{1}{2} |\langle \sqrt{1 - \alpha p^2} \rangle |$ or generalized commutation relation $[q, p] = i \sqrt{1 - \alpha p^2},$ one can still  consider the zero-point energy as a volume energy attributed to  the first term appearing within the square root term in $\frac{1}{2} |\langle \sqrt{1 - \alpha p^2} \rangle |$ (or in $ i \sqrt{1 - \alpha p^2}$). However, we have already shown that the second term within the square root terms, having contributions of deformation parameter $\alpha$, corresponds to a holographic surface energy density.
Therefore, one may interpret the first and second terms ``1" and $``\alpha p^2$ within the square root term, corresponding to the bare volume energy density and physical surface energy density, respectively. Now, the solution of cosmological constant problem lies
in the fact that the bare cosmological constant of order $M_P^4$ coming from
the volume energy density associated to ``1" in the Generalized Uncertainty Principle does not contribute to the holographic based physical cosmological constant $\Lambda \sim \rho_{_{\alpha}}$ coming from the deformation parameter
$\alpha$. 

In conclusion, the physical cosmological constant which is coupled to
the Einstein equations comes just from the holographic surface energy density
originating from {\it ``quantum gravitational modifications included in GUP''}, and the bare cosmological constant as a volume energy density receiving
huge contributions from {\it ``quantum gravitational corrections to zero pint energy''} is naturally ruled out from being coupled to the Einstein equations.

For the case of $\rho_{_{\alpha}}(\omega)$,  the holographic energy density depends  on both the quantum
deformation
parameter $\alpha$ and the classical equation of state
parameter $\omega$. For any value of $\omega$ given by (\ref{www}), the radius of Einstein static universe and the holographic
energy density  are determined by the equation (\ref{20}). Moreover, according to the equation (\ref{20'}), the variable holographic energy density $\rho_{_{\alpha}}(\omega)$ is proportional to the constant holographic energy density $\rho_{_{\alpha}}$.
Therefore, one may write $\rho_{_{\alpha}}(\omega)={\rho_{_{\alpha}}}/{4(1-3w)}=\rho_{_{\alpha}}^{eff}$ where $\rho_{_{\alpha}}^{eff}$ is considered as {\it effective} cosmological
constant. Thus, the same discussion on the holographic origin of the observed
cosmological
constant and the solution of cosmological constant problem can be applied to this case, as well.

\section*{Summary and Conclusion}
We have studied the Einstein Static Universe in the framework of Generalized Uncertainty Principle. We have shown that the deformation parameter can induce an energy density subject to GUP which
obeys the holographic feature and plays the role of a cosmological constant. Using the holographic feature of  GUP energy density, we have introduced natural IR-UV cut-offs which are dependent on the effective equation of state parameter
subject to GUP. Moreover, we have realized that the Einstein equations are naturally
coupled to the tiny holographic surface
energy density (interpreted as physical cosmological constant) instead of large volume energy density (interpreted as bare cosmological constant). Then, we have proposed a solution
to the cosmological constant problem.

\section*{Acknowledgments}
This work has been supported financially by a grant number 217/D/17739 from Azarbaijan Shahid Madani University.

\end{document}